\documentclass[12pt,a4paper,twoside,epsf]{article}
\usepackage{epsfig}
\usepackage{baltlat1}
\usepackage{wrapfig}
\begin{document}
\ \
\vspace{0.5mm}

\setcounter{page}{1}
\vspace{8mm}

\titlehead{Baltic Astronomy, vol.12, XXX--XXX, 2003.}

\titleb{EVIDENCE FOR TOROIDAL B-FIELD STRUCTURES IN BL~LAC OBJECTS}

\begin{authorl}
\authorb{D.C.~Gabuzda}{ }
\authorb{\'E.~Murray}{ }
\authorb{P.~Cronin}{ }
\end{authorl}

\begin{addressl}
\addressb{ }{Department of Physics, University College Cork, Cork, Ireland}
\end{addressl}

\submitb{Received: 34 October 2003}

\begin{abstract}
If the transverse jet {\bf B} fields observed in BL~Lac objects on parsec 
scales represent the toroidal component of the jet {\bf B} fields, 
this should give rise to rotation-measure gradients {\em across} the
jets, due to the systematic change in the line-of-sight component
of the jet {\bf B} field. We have found evidence for such rotation-measure 
gradients in one-third to one-half of the 1~Jy BL~Lac objects we have 
analyzed. We present and discuss these new results, together with some 
of their implications for our understanding of the parsec-scale jets of AGN.

\end{abstract}


\resthead{Toroidal B-Field Structures}{D.C.~Gabuzda, \'E.~Murray, P.Cronin}



\sectionb{1}{INTRODUCTION}

BL~Lac objects are highly variable, appreciably polarized Active Galactic
Nuclei that are observationally similar to radio-loud quasars in many 
respects, but display systematically weaker optical line emission.  
VLBI polarization observations of radio-loud BL~Lac objects have shown
a tendency for the dominant magnetic ({\bf B}) fields in the parsec-scale
jets to be transverse to the local jet direction (Gabuzda, Pushkarev, \&
Cawthorne 2000 \& references therein). This has often been interpreted as
evidence for relativistic shocks that enhance the {\bf B}-field component
in the plane of compression, perpendicular to the direction of propagation
of the shock (Laing 1980; Hughes, Aller, \& Aller 1989).

It has been suggested more recently that the transverse jet {\bf B}
fields of BL~Lac objects often correspond to the toroidal {\bf B}-field
component of the jet itself (e.g., Gabuzda 1999, 2003; Gabuzda \& Pushkarev
2002). Such fields could come about as a result of the  ``winding up'' of 
an initial ``seed'' field with a significant longitudinal component by the 
rotation of the central accreting object (e.g.  Nakamura, Uchida, \& Hirose 
2001; Lovelace et al.  2002; Hujeirat et al. 2003; Lynden-Bell 2003; 
Tsinganos \& Bogovalov 2002).
It is therefore of interest to identify robust
observational tests that can distinguish between transverse {\bf B}
fields due to a toroidal field component and due to shock compression.
One possibility is to search for rotation-measure (RM) gradients {\em across}
the jets, which should arise in the case of a toroidal {\bf B}-field
structure due to the systematic change in the line-of-sight magnetic
field across the jet.  Asada et al. (2002) claim to have detected such
a gradient across the VLBI jet of 3C273.  We present and discuss here
evidence for transverse RM gradients in a number of BL~Lac objects.

\sectionb{2}{OBSERVATIONS AND RESULTS}

\begin{wrapfigure}{i}[0pt]{72mm}
\centerline{\psfig{figure=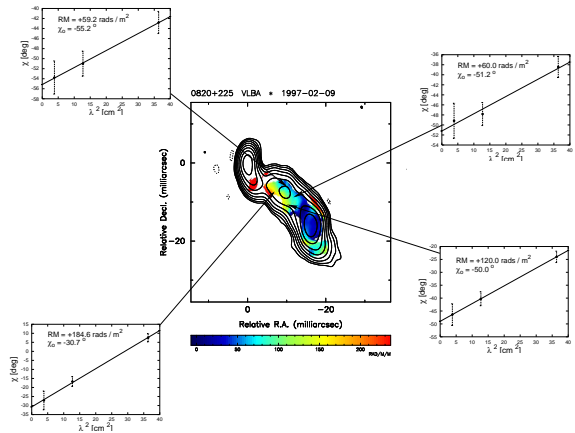,width=65truemm,angle=-90,clip=}}
\caption{6-cm $I$ contours of 0820+225 with the RM distribution
superposed in grey scale and plots of $\chi$ versus $\lambda^2$. 
}
\end{wrapfigure}

Our analysis was based on polarization observations of the 34 BL~Lac objects 
in the complete sample defined by K\"uhr \& Schmidt (1990) carried out in 
February 1997 (1997.11, 24 sources) and June 1999 (1999.46, six sources) 
at 6, 4 \& 2~cm using the NRAO Very Long Baseline Array.  After eliminating 
sources that did not clearly show polarization outside the core at all three 
wavelengths, we were left with 22 sources suitable for rotation-measure 
mapping.  A brief description of the data analysis is given by Gabuzda \& 
Murray (2003), and more complete information will be given in a future paper.  

We found clear RM gradients transverse to the VLBI jets in six of
the objects, probable RM gradients in two more sources, 
and tentative RM gradients in three more. 
The figure shows the 6-cm VLBI $I$ image of 0820+25 with its parsec-scale
RM distributions superposed in grey scale. The
accompanying plots show the observed polarization position angles,
$\chi$, as a function of the observing wavelength squared, $\lambda^2$,
for individual locations in the VLBI jet. In all cases, the observed
$\chi$ values are in good agreement
with the $\lambda^2$ law expected for Faraday rotation.

\sectionb{3}{DISCUSSION}


It is natural to interpret the observed transverse RM gradients as
reflecting a toroidal or helical {\bf B} field associated with these 
VLBI jets. In this case, the gradients are due to the systematic change 
in the line-of-sight {\bf B} field component across the jet. The fact that 
we detected transverse RM gradients in one-third to one-half of the 
sample sources studied suggests that they are quite common in BL~Lac 
objects. In two cases for which RM maps are available at multiple epochs
(Mrk501: Croke et al., these proceedings; 1803+784: Zavala \& Taylor 2003),
the independent RM maps show transverse RM gradients in the same sense,
providing evidence that the phenomenon is stable over time scales of
one to several years. 

A helical {\bf B}-field could come about via the ``winding up'' of
a seed field threading the central accretion disk by the joint action of
the rotation of the accretion disk and the jet outflow (e.g., Nakamura, 
Uchida, \& Hirose 2001; Lovelace et al.  2002; Hujeirat et al. 2003; 
Lynden-Bell 2003; Tsinganos \& Bogovalov 2002).  It is intriguing that
a {\bf B} field with a predominant toroidal component would also come
about if a non-zero current flows along the jet (e.g., Pariev, Istomin \&
Beresnyak 2003; Lyutikov 2003).
Indeed, if we are seeing a dominant toroidal {\bf B}-field component,
basic physics demands that there {\em must} be currents
flowing in the region enclosed by the field.

If we view a toroidal or helical {\bf B} field
at $90^{\circ}$ to the jet axis in the rest
frame of the source and the distribution of thermal electrons is
approximately uniform, we would expect to observe a rotation measure
close to zero along the jet axis (since the line-of-sight {\bf B}-field
component there is close to zero) and RMs of opposite sign on either side
of the axis.  When the {\bf B} field is viewed at some other angle to the 
jet axis, there will still be a systematic gradient in the RM across the 
jet, however the gradient ``peak'' will be shifted, and the RM will not 
necessarily pass through zero. Note that photons emitted at $90^{\circ}$
to the jet axis in the source frame will be observed in the observer's
frame to propagate at an angle $\theta = 1/\gamma$ to the jet axis, where
$\gamma$ is the Lorentz factor for the jet's motion. 

The results of our search for rotation-measure gradients transverse
to the VLBI jets of BL Lac objects yield firm evidence for such
gradients in several sources, lending strong support to earlier arguments
that the ``transverse'' {\bf B} fields that are often observed in these 
objects are associated with toroidal or helical structure of the intrinsic
jet {\bf B} fields. This underlines the view of these jets as fundamentally
electromagnetic structures, and suggests that they may well carry non-zero
currents. It also suggests that the jets may be {\em Poynting-flux
dominated}, which is of cardinal theoretical importance.

A version of this paper with a colour figure can be found at astro-ph/0309668.
\goodbreak

\References
\ref
Asada~K. et al. 2002, PASJ, 54, L39
\ref
Gabuzda~D.~C. 1999, New Astronomy Reviews, 43, 695
\ref
Gabuzda~D.~C. 2003, New Astronomy Reviews, 47, 599 
\ref
Gabuzda~D.~C., Murray~\'E. 2003, astro-ph/0309668
\ref
Gabuzda~D.~C., Pushkarev~A.~B. 2002 in {\smalit Particles
and Fields in Radio Galaxies}, ASP Conf.\ Ser.\ 250, eds. R. Laing and 
K. Blundell,\ p.\,180
\ref
Gabuzda~D.~C., Pushkarev~A.~B., Cawthorne~T.~V. 2000, MNRAS, 319, 1109
\ref
Hugeirat~A., Livio~M., Camenzind~M., Burkert~A. 2003, A\&A, 408, 415
\ref
Hughes~P.~A., Aller~H.~D., Aller~M.~A. 1989, ApJ, 341, 68
\ref
Laing~R. 1980, MNRAS, 193, 439
\ref
Lovelace~R.~V.~E., Li~H., Koldoba~A.~V., Ustyugova~G.~V., Romanova~M.~M.
2002, ApJ, 572, 445
\ref
Lynden-Bell~D. 2003, MNRAS, 341, 1360
\ref
Lyutikov~M.  2003, New Astronomy Reviews, 47, 513
\ref
Nakamura~M., Uchida~Y., Hirose~S. 2001, New Astronomy, 6, 61
\ref
Pariev~V.~I., Istomin~Ya.~N., Beresnyak~A.~R. 2003,
A\& A, 403, 805
\ref 
Tsinganos~K., Bogovalov~S. 2002, MNRAS, 337, 553
\ref
Zavala~R.~T., Taylor~G.~B. 2003, ApJ, 589, 126

\end{document}